\documentclass[aps,prl,reprint,superscriptaddress]{revtex4-1}
\usepackage{amsmath}
\usepackage{amsfonts}
\usepackage{amssymb}
\usepackage{amsthm}
\usepackage{graphicx}
\usepackage{CJK}
\usepackage{color}
\usepackage{times}
\usepackage{mathptmx}
\usepackage[colorlinks, linkcolor=blue, urlcolor=blue, anchorcolor=blue, citecolor=blue]{hyperref}
\usepackage{mathtools}

\begin{document}


\title{Ultra-high differential mobility and velocity of N\'{e}el walls in spin valves 
	with planar-transverse polarizers under low perpendicularly injected currents}


\author{Mei Li}
\affiliation{Physics Department, Shijiazhuang University, Shijiazhuang, Hebei 050035, China}
\author{Zhong An}
\affiliation{College of Physics and Information Engineering, Hebei Advanced Thin Films Laboratory, Hebei Normal University, Shijiazhuang 050024, China}
\author{Jie Lu}
\email{jlu@hebtu.edu.cn}
\affiliation{College of Physics and Information Engineering, Hebei Advanced Thin Films Laboratory, Hebei Normal University, Shijiazhuang 050024, China}


\date{\today}

\begin{abstract}
Transverse domain wall (TDW) dynamics in long and narrow spin valves with perpendicular
current injection is theoretically investigated. 
We demonstrate that stable traveling-wave motion of TDWs with finite velocity survives for strong enough
planar-transverse polarizers.
For typical ferromagnetic materials (for example, Co) and achievable spin polarization ($P=0.6$), 
TDWs acquire a velocity of $10^3$ m/s under a current density below $10^7$ $\mathrm{A/cm^2}$.
This efficiency is comparable with that of perpendicular polarizers. 
More importantly, in this case the wall has ultra-high ``differential mobility"
around the onset of stable wall excitation.
Our results open new possibilities for developing magnetic nanodevices based on TDW propagation with low energy consumption.
Also, analytics for parallel and perpendicular polarizers perfectly explains existing simulation findings.
Finally, further boosting of TDWs by external uniform transverse magnetic fields is
investigated and turns out to be efficient.
\end{abstract}


\maketitle

\section{\label{Section_introduction} I. Introduction} 
Tremendous progress in fabrication technology of non-volatile magnetic nanodevices
has led to a great revolution in modern information industry\cite{Leeuw_RPP_1980,Bauer_RMP_2005,Klaui_JPCM_2008}.
In these nanodevices, magnetic domains with different orientations build zeros and ones in binary world.
Intermediate regions separating these domains are the domain walls (DWs) and their motion leads to 
the data transformation\cite{XiongG_2005_Science,Parkin_2008_Science,Koopmans_2012_nanotech,Thomas_JAP_2012,Parkin_2015_nanotech,jlu_NJP_2019}.
Generally, DWs' motion can be induced by magnetic fields, spin-polarized currents or temperature gradient, etc.
Among them, the current-induced case is the easiest to implement in real experiments.

Historically, the earliest current-induced driving mechanism of DWs is the spin-transfer torque (STT). 
It was first calculated in a magnetic multilayer, in which two ferromagnetic (FM)
layers are single-domained with ``current perpendicular to the plane (CPP)" configuration\cite{Slonczewski_JMMM_1996}.
The resultant STT is the so-called Slonczewski torque (SLT) and proportional to
$\mathbf{m}\times(\mathbf{m}\times\mathbf{m}_{\mathrm{p}})$ in which $\mathbf{m}$ and $\mathbf{m}_{\mathrm{p}}$
are normalized magnetization vectors in the thin (free) and thick (pinned) layers.
Meantime, another torque ($\propto \mathbf{m}\times\mathbf{m}_{\mathrm{p}}$) also exists and is
usually referred as the field-like torque (FLT) since now $\mathbf{m}_{\mathrm{p}}$ acts like an effective field.
Later in magnetic nanostrips with currents flowing in strip plane (CIP), adiabatic and nonadiabatic STTs
are proposed and can be viewed as the
continuous limits of SLT and FLT, respectively\cite{Zhang_PRL_2004,Tatara_PhysRep_2008}.
The adiabatic STT induces the initial DW movement but the final steady wall velocity
is determined by the nonadiabatic STT. However, since the exchange interaction avoids abrupt variation
of magnetization, CIP current densities of several $10^{8}\ \mathrm{A/cm^2}$ only induce DW velocity around $100$ m/s.

To increase current efficiency, long and narrow spin valves (LNSVs) or
magnetic tunneling junctions (MTJs) with CPP configuration are proposed to be host systems\cite{Fert_JAP_2002,Fert_APL_2003,Lim_APL_2004}.
In these multilayers, DWs in free layers are driven to move along the long axis by
spin-polarized current filtered by pinned layers (polarizers).
Early simulations on parallel and perpendicular polarizers only considered SLTs 
and asserted that the current efficiency can not be increased too much\cite{Rebei_Mryasov_PRB_2006,Kawabata_IEEE_2011}.
In 2009, a significant breakthrough\cite{Khvalkovskiy_PRL_2009} was made by Khvalkovskiy \textit{et. al.} 
in which numerical simulations with both SLT and FLT
revealed that to achieve a DW velocity of
100 m/s, the CPP current density for parallel polarizers is lowered to $10^{7}\ \mathrm{A/cm^2}$,
while for perpendicular polarizers, the CPP current density is further decreased to $10^{6}\ \mathrm{A/cm^2}$.

Later, two series of experimental works were carried out.
First, in LNSVs\cite{Boone_PRL_2010_exp} and half-ring MTJs\cite{Grollier_NatPhys_2011,Metaxas_SciRep_2013,Grollier_APL_2013} 
with CPP configuration, transport measurements confirm
that DWs can propagate with velocities as high as 500-800 m/s at current density below $10^{7}\ \mathrm{A/cm^2}$.
Second, in ZigZag LNSVs with CIP configuration high DW velocities (150-600 m/s) are obtained for current
densities of $(2\sim5)\times 10^{7}\ \mathrm{A/cm^2}$ by using photoemission electron microscopy combined 
with X-ray magnetic circular dichroism\cite{Pizzini_APE_2009,Pizzini_PRB_2010,Pizzini_PRB_2011,Pizzini_JPCM_2012}. 
Vertical spin current coming with
spin flux transformation from pinned layers to free layers via spacers (thus similar to CPP) is
suggested to provide a potential explanation for this velocity boosting.

Except for these concentrated explorations on parallel and perpendicular polarizers,
LNSVs with planar-transverse polarizers have not received enough attention in existing literatures.
Within a mature Lagrangian framework\cite{He_EPJB_2013}, in this work we show that 
stable traveling-wave motion of DWs with finite velocity exists for strong enough planar-transverse polarizers.
The resulting current efficiency is comparable with that of perpendicular polarizers. 
Furthermore, ultra-high ``differential mobility" emerges around the onset of stable wall excitation.
Also, we provide analytics for parallel and perpendicular polarizers which perfectly explains existing simulations.
At last, further boosting of DWs by uniform transverse magnetic fields (UTMFs) are
studied with the help of one-dimensional asymptotic expansion method (1D-AEM)\cite{Goussev_PRB_2013,Goussev_Royal_2013,jlu_PRB_2016,jlu_SciRep_2017,jlu_Nanomaterials_2019}.

\section{\label{Section_Lagrangian} II. Model and method}
\begin{figure} [htbp]
	\centering
	\includegraphics[width=0.45\textwidth]{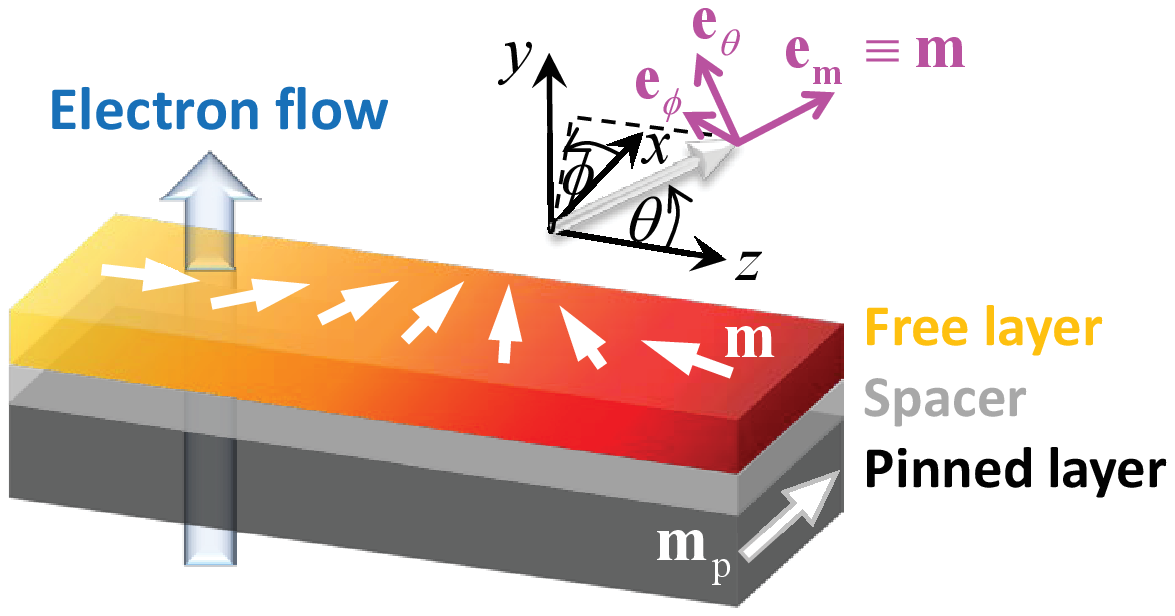}
	\caption{(Color online) Sketch of a LNSV with CPP configuration, which is a three-layer structure: 
		a pinned FM layer ($\mathbf{m}_{\mathrm{p}}$, polarizer), a NM metallic spacer and a free FM layer ($\mathbf{m}$).
		A DW in the free layer is driven to move along the long axis of LNSV by	perpendicularly injected currents.
		($\mathbf{e}_x,\mathbf{e}_y,\mathbf{e}_z$) is the global Cartesian coordinate system, and
		($\mathbf{e}_{\mathbf{m}},\mathbf{e}_{\theta},\mathbf{e}_{\phi}$) forms the local spherical coordinate
		system associated with $\mathbf{m}$.}\label{fig1}
\end{figure}

We consider a LNSV with CPP configuration (see Fig. \ref{fig1}), which is composed of 
three layers: a free FM layer with tunable magnetization texture,
a nonmagnetic (NM) metallic spacer and a pinned FM layer with a fixed magnetization orientation (polarizer).
The global Cartesian coordinate system is as follows:
$\mathbf{e}_z$ is along the long axis of LNSV, $\mathbf{e}_y$ follows the electron flow direction
(from pinned to free layer) and $\mathbf{e}_x=\mathbf{e}_y\times\mathbf{e}_z$.
The polarizer is usually made of hard ferromagnetic materials. Its magnetization ($\mathbf{m}_{\mathrm{p}}$) 
has three typical choices:
(a) $\mathbf{m}_{\mathrm{p}}=\mathbf{e}_z$ (parallel),
(b) $\mathbf{m}_{\mathrm{p}}=\mathbf{e}_y$ (perpendicular) and
(c) $\mathbf{m}_{\mathrm{p}}=\mathbf{e}_x$ (planar-transverse).
Electrons flow from the polarizer to the free layer via the metallic spacer with
density $J_e(>0)$. Thus the charge current is $J_{\mathrm{charge}}=-J_e\mathbf{e}_y$.

The normalized magnetization $\mathbf{m}$ of the free layer can be fully
described by its polar angle $\theta$ and azimuthal angle $\phi$.
The associated local spherical coordinate system is denoted as
($\mathbf{e}_{\mathbf{m}},\mathbf{e}_{\theta},\mathbf{e}_{\phi}$).
Then $\mathbf{m}_{\mathrm{p}}$ is decomposed into
\begin{equation}\label{mp_expression_in_sperical}
\mathbf{m}_{\mathrm{p}}=p_{\mathbf{m}}\mathbf{e}_{\mathbf{m}}+p_{\theta}\mathbf{e}_{\theta}+p_{\phi}\mathbf{e}_{\phi},
\end{equation}
with
\begin{equation}\label{mp_expression_in_sperical_definitions}
\begin{split}
p_{\mathbf{m}}&=\sin\theta_{\mathrm{p}}\cos(\phi-\phi_{\mathrm{p}})\sin\theta+\cos\theta_{\mathrm{p}}\cos\theta,   \\
p_{\theta}&=\sin\theta_{\mathrm{p}}\cos(\phi-\phi_{\mathrm{p}})\cos\theta-\cos\theta_{\mathrm{p}}\sin\theta,   \\
p_{\phi}&=-\sin\theta_{\mathrm{p}}\sin(\phi-\phi_{\mathrm{p}}),
\end{split}
\end{equation}
where $\theta_{\mathrm{p}}$ ($\phi_{\mathrm{p}}$) is the polar (azimuthal) angle of $\mathbf{m}_{\mathrm{p}}$.

The magnetic energy of the free layer includes
the exchange, crystalline anisotropy, magnetostatic and FLT-induced effective potential.
Following He's work\cite{He_EPJB_2013}, we have
\begin{equation}\label{magnetic_energy_density_total_HePB}
\mathcal{E}[\mathbf{m}]=\mathcal{E}_0[\mathbf{m}]-\mu_0 M_s^2 \xi_{\mathrm{CPP}} \frac{J_e}{J_{\mathrm{p}}}\frac{b_{\mathrm{p}}}{c_{\mathrm{p}}}\ln(1+c_{\mathrm{p}} p_{\mathbf{m}}),
\end{equation}
with
\begin{equation}\label{magnetic_energy_density_0}
\mathcal{E}_0[\mathbf{m}]=J\left(\frac{\partial\mathbf{m}}{\partial z}\right)^2+\mu_0 M_s^2\left(-\frac{1}{2}k_{\mathrm{E}}m_z^2+\frac{1}{2}k_{\mathrm{H}}m_y^2\right),
\end{equation}
in which the magnetostatic energy has been described by local
quadratic terms of $M_{x,y,z}$ by means of three average demagnetization factors\cite{jlu_PRB_2016}.
$J$ is the exchange stiffness, $\mu_0$ is the vacuum permeability,
$\xi_{\mathrm{CPP}}$ describes the relative strength of FLT over SLT,
$k_{\mathrm{E}}(k_{\mathrm{H}})$ is the total anisotropy coefficient
along the easy (hard) axis of the free layer and $M_s$ is the saturation magnetization.
In addition, $J_{\mathrm{p}}\equiv 2 \mu_0 e d M_s^2/\hbar$
where $d$ is the thickness of free layer, $e(>0)$ is the absolute charge of electron
and $P$ is the spin polarization of the current.
At last, the two dimensionless parameters $b_{\mathrm{p}}=4P^{3/2}/[3(1+P)^3-16P^{3/2}]$ 
and $c_{\mathrm{p}}=(1+P)^3/[3(1+P)^3-16P^{3/2}]$ reproduce Slonczewski's original spin polarization factor
$g\equiv [-4+(1+P)^3(3+\mathbf{m}\cdot\mathbf{m}_{\mathrm{p}})/(4P^{3/2})]^{-1}$\cite{Slonczewski_JMMM_1996}
by $g=b_{\mathrm{p}}/(1+c_{\mathrm{p}} p_{\mathbf{m}})$.

The magnetization dynamics in the free layer is described by the Lagrangian
$L=\int \mathcal{L} \mathrm{d}^3\mathbf{r}$ with density
\begin{equation}\label{Lagrangian_density}
\mathcal{L}=\frac{\mu_0 M_s}{\gamma_0}\dot{\phi}\cdot (1-\cos\theta)-\mathcal{E},
\end{equation}
in which $\gamma_0=\mu_0\gamma$ with $\gamma$ being the gyromagnetic ratio and a dot means $\partial/\partial t$.
To include the Gilbert damping and the SLT-induced anti-damping processes,
an extra dissipation functional $F=\int \mathcal{F} \mathrm{d}^3\mathbf{r}$ is introduced with density
\begin{equation}\label{Dissipation_density_HePB}
\frac{\mathcal{F}}{\mu_0 M_s^2}=\frac{\alpha}{2}\frac{\dot{\theta}^2+\dot{\phi}^2\sin^2\theta}{\gamma_0 M_s}-g \frac{J_e}{J_{\mathrm{p}}} (p_{\theta}\sin\theta\dot{\phi}-p_{\phi}\dot{\theta}).
\end{equation}
The corresponding generalized Eular-Lagrangian equation
\begin{equation}\label{Eular_Lagrangian_equation}
\frac{\mathrm{d}}{\mathrm{d}t}\left(\frac{\delta\mathcal{L}}{\delta\dot{X}}\right)-\frac{\delta\mathcal{L}}{\delta X}+\frac{\delta\mathcal{F}}{\delta \dot{X}}=0,
\end{equation}
provides dynamical descriptions of TDWs, where $X$ represents related collective coordinates.

Early simulations confirmed that in FM nanostrips with small enough cross section,
transverse DWs (TDWs) have the lowest energy among all meta-stable states\cite{McMichael_IEEE_1997,Thiaville_JMMM_2005}.
In 2012, further simulations revealed that the stability range of TDW in free layers of LNSVs
can be shifted towards larger cross section compared with monolayer strips, due to a magnetostatic
screening effect between the free and pinned layers\cite{Pizzini_APL_2012}.
Therefore the configuration space of DWs in this work is the TDW with generalized Walker profile\cite{Walker_JAP_1974}
\begin{equation}\label{Walker_static_generalized}
\ln\tan\frac{\vartheta(z,t)}{2}=\eta\frac{z-q(t)}{\Delta(t)},\quad \phi(z,t)\equiv\varphi(t),
\end{equation}
in which $\eta=+1$ or $-1$ represents head-to-head (HH) or tail-to-tail (TT) TDWs, respectively.
Note that in many 1D collective-coordinate analysis, the tilting angle $\varphi(t)$ and 
wall center position $q(t)$ [or wall velocity $\dot{q}(t)$] are the two collective coordinates 
meanwhile assuming fixed wall width $\Delta(t)$\cite{Boone_PRL_2010_theo,Tatara_JPDAP_2011}. 
However, the wall width does change considerably as the wall tilting angle varies 
if the material is magnetically biaxial. 
Even for uniaxial materials, the strip geometry will induce an effective hard axis 
in the normal direction perpendicular to strip plane. 
Based on these facts, we therefore view the wall width as the third collective coordinate.

In Eq. (\ref{Eular_Lagrangian_equation}), by letting $X$ take $q(t)$, $\varphi(t)$, $\Delta(t)$ successively,
and integrating over the long axis (i.e. $\int_{-\infty}^{+\infty}\mathrm{d}z$), we obtain the following
dynamic equations
\begin{subequations}\label{Dynamical_equation_original_HePB}
	\begin{align}
	\frac{\dot{\varphi}+\alpha\eta\dot{q}/\Delta}{\gamma_0 M_s}&=b_{\mathrm{p}}\frac{J_e}{J_{\mathrm{p}}}\left[ p_{\varphi} U(\varphi)-\frac{\xi_{\mathrm{CPP}}}{2c_{\mathrm{p}}}\ln\frac{1-c_{\mathrm{p}} \cos\theta_{\mathrm{p}}}{1+c_{\mathrm{p}}\cos\theta_{\mathrm{p}}} \right], \\
	\frac{\alpha\dot{\varphi}-\eta\dot{q}/\Delta}{\gamma_0 M_s} &= b_{\mathrm{p}}\frac{J_e}{J_{\mathrm{p}}}\left[\xi_{\mathrm{CPP}} p_{\varphi} U(\varphi)+\frac{1}{2c_{\mathrm{p}}}\ln\frac{1-c_{\mathrm{p}}\cos\theta_{\mathrm{p}}}{1+c_{\mathrm{p}}\cos\theta_{\mathrm{p}}} \right] \nonumber \\
	& \qquad -k_{\mathrm{H}}\sin\varphi\cos\varphi, \\
	\frac{\pi^2\alpha}{6\gamma_0 M_s}\frac{\dot{\Delta}}{\Delta}&= b_{\mathrm{p}}\frac{J_e}{J_{\mathrm{p}}}\left[\xi_{\mathrm{CPP}}W(\varphi)-p_{\varphi} U(\varphi)\ln\frac{1-c_{\mathrm{p}}\cos\theta_{\mathrm{p}}}{1+c_{\mathrm{p}}\cos\theta_{\mathrm{p}}} \right] \nonumber  \\
	& \qquad +\left(\frac{l_0^2}{\Delta^2}-k_{\mathrm{E}}-k_{\mathrm{H}}\sin^2\varphi\right).
	\end{align}
\end{subequations}
with
\begin{equation}\label{UWPsi_definition_Wnew}
\begin{split}
U(\varphi)&\equiv\chi/\sqrt{1-c_{\mathrm{p}}^2\left[\sin^2\theta_{\mathrm{p}}\cos^2(\varphi-\phi_{\mathrm{p}})+\cos^2\theta_{\mathrm{p}}\right]},  \\
W(\varphi)&\equiv \frac{1}{2 c_{\mathrm{p}}}\left[\frac{\pi^2}{4}+\frac{1}{4}\ln^2\frac{1-c_{\mathrm{p}}\cos\theta_{\mathrm{p}}}{1+c_{\mathrm{p}}\cos\theta_{\mathrm{p}}}-\chi^2\right],  \\
\chi&\equiv\arccos\frac{c_{\mathrm{p}}\sin\theta_{\mathrm{p}}\cos(\varphi-\phi_{\mathrm{p}})}{\sqrt{1-c_{\mathrm{p}}^2\cos^2\theta_{\mathrm{p}}}},
\end{split}
\end{equation}
and  $l_0\equiv\sqrt{2J/(\mu_0 M_s^2)}$ being the exchange length of the free layer.
Note that in the definition of function $W(\varphi)$ in Eq. (\ref{UWPsi_definition_Wnew}),
our calculation supports an additional ``1/2" factor compared with He's original work.

\section{III. DW dynamics under planar-transverse polarizers}
For planar-transverse polarizers, $\theta_{\mathrm{p}}=\pi/2$ and $\phi_{\mathrm{p}}=0$.
The dynamical equations evolve to
\begin{subequations}\label{Dynamical_equation_pt_polarizer}
	\begin{align}
	\frac{1+\alpha^2}{\gamma_0 M_s\sin\varphi}\frac{\eta\dot{q}}{\Delta}&=\left[k_{\mathrm{H}}\cos\varphi-(\alpha-\xi_{\mathrm{CPP}})b_{\mathrm{p}}\frac{J_e}{J_{\mathrm{p}}}\widetilde{U}(\varphi)\right],   \\
	\frac{1+\alpha^2}{\gamma_0 M_s\sin\varphi}\dot{\varphi}&=-\left[(1+\alpha\xi_{\mathrm{CPP}})b_{\mathrm{p}}\frac{J_e}{J_{\mathrm{p}}}\widetilde{U}(\varphi)+\alpha k_{\mathrm{H}}\cos\varphi\right],  \\
	\frac{\pi^2\alpha}{6\gamma_0 M_s}\frac{\dot{\Delta}}{\Delta}&=\left(\frac{l_0^2}{\Delta^2}-k_{\mathrm{E}}-k_{\mathrm{H}}\sin^2\varphi\right)+\xi_{\mathrm{CPP}}b_{\mathrm{p}}\frac{J_e}{J_{\mathrm{p}}}\widetilde{W}(\varphi),
	\end{align}
\end{subequations}
in which
\begin{equation}\label{UWPsi1_definition}
\begin{split}
\widetilde{U}(\varphi)&=\frac{\tilde{\chi}}{\sqrt{1-c_{\mathrm{p}}^2\cos^2\varphi}},\quad \widetilde{W}(\varphi)=\frac{1}{2c_{\mathrm{p}}}\left(\frac{\pi^2}{4}-\tilde{\chi}^2\right),   \\
\tilde{\chi}&=\arccos(c_{\mathrm{p}}\cos\varphi).
\end{split}
\end{equation}

For steady traveling-wave mode, $\dot{\varphi}=0$ and $\dot{\Delta}=0$.
This leads to two branches of solution:
\begin{equation}\label{Solution_branch_1_pt_polarizer}
\begin{split}
\varphi_0&=n\pi,\quad v_0=0,  \\ \Delta(\varphi_0)&=l_0\left[k_{\mathrm{E}}-\xi_{\mathrm{CPP}}b_{\mathrm{p}}\frac{J_e}{J_{\mathrm{p}}}\widetilde{W}(\varphi_0)\right]^{-1/2},
\end{split}
\end{equation}
and
\begin{equation}\label{Solution_branch_2_pt_polarizer}
\begin{split}
\cos\varphi'_0&=-\frac{1+\alpha\xi_{\mathrm{CPP}}}{\alpha k_{\mathrm{H}}}b_{\mathrm{p}}\frac{J_e}{J_{\mathrm{p}}}\widetilde{U}(\varphi'_0),   \\
v'_0&=\frac{\eta\Delta(\varphi'_0)\gamma_0 k_{\mathrm{H}} M_s}{1+\alpha\xi_{\mathrm{CPP}}}\sin\varphi'_0\cos\varphi'_0,    \\
\Delta(\varphi'_0)&=l_0\left[k_{\mathrm{E}}+k_{\mathrm{H}}\sin^2\varphi'_0-\xi_{\mathrm{CPP}}b_{\mathrm{p}}\frac{J_e}{J_{\mathrm{p}}}\widetilde{W}(\varphi'_0)\right]^{-\frac{1}{2}}.
\end{split}
\end{equation}

\begin{figure} [htbp]
	\centering
	\includegraphics[width=0.48\textwidth]{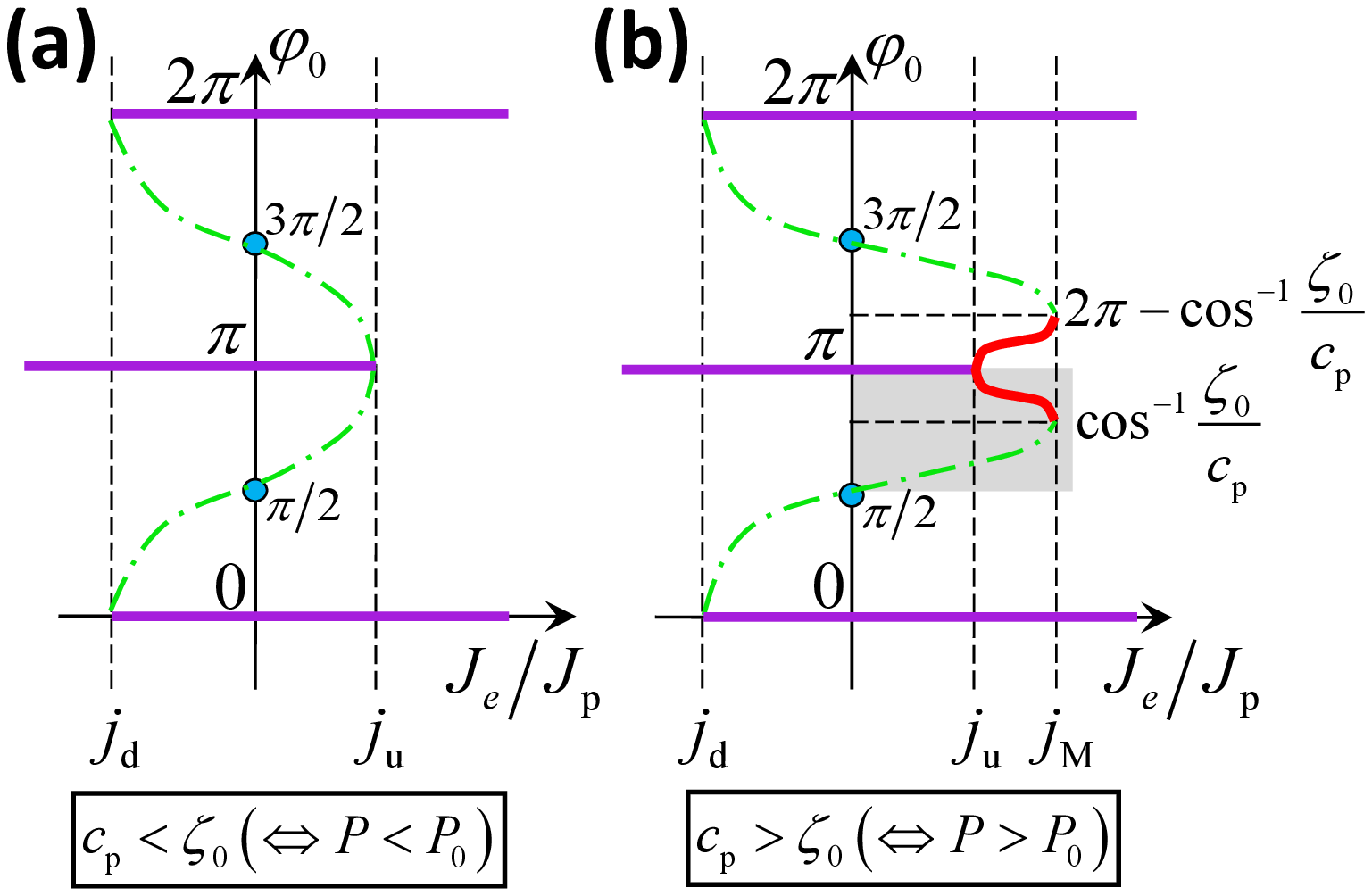}
	\caption{(Color online) Illustration of the solution branch
		in Eq. (\ref{Solution_branch_2_pt_polarizer}): (a) $c_{\mathrm{p}}<\zeta_0$,
		(b) $c_{\mathrm{p}}>\zeta_0$. In both cases, violet solid lines represent
		the stable solution branch in Eq. (\ref{Solution_branch_1_pt_polarizer}) with zero velocity,
		and green dash-dot curves represent the unstable part of solution branch
		in Eq. (\ref{Solution_branch_2_pt_polarizer}). In addition, red solid curves
		in (b) indicate the stable part of solution branch
		in Eq. (\ref{Solution_branch_2_pt_polarizer}). 
		The shaded area in (b) will be calculated in details in Fig. \ref{fig3}. }\label{fig2}
\end{figure}

For the first branch in Eq. (\ref{Solution_branch_1_pt_polarizer}),
For the variation $\varphi=\varphi_0+\delta\varphi$,
Eq. (\ref{Dynamical_equation_pt_polarizer}b) provides
\begin{equation}\label{Stability_analysis_phi_pt_polarizer}
\begin{split}
\frac{\partial(\ln\delta\varphi)}{\partial t}&=-\frac{\gamma_0 M_s}{1+\alpha^2}\bigg\{(-1)^n(1+\alpha\xi_{\mathrm{CPP}})b_{\mathrm{p}}\frac{J_e}{J_{\mathrm{p}}}\times   \\
& \quad (1-c_{\mathrm{p}}^2)^{-1/2}\arccos\left[(-1)^n c_{\mathrm{p}}\right]+\alpha k_{\mathrm{H}}\bigg\}.
\end{split}
\end{equation}
The stability of $\varphi_0-$solution requires the terms in curly braces to be positive. This leads to $J_e/J_{\mathrm{p}}>j_{\mathrm{d}}$ ($n$ is even) or
$J_e/J_{\mathrm{p}}<j_{\mathrm{u}}$ ($n$ is odd), where
\begin{equation}\label{jdown_jup_definitions}
\begin{split}
j_{\mathrm{u}}&\equiv\frac{\alpha k_{\mathrm{H}}}{1+\alpha\xi_{\mathrm{CPP}}}\cdot\frac{\sqrt{1-c_{\mathrm{p}}^2}}{b_{\mathrm{p}}\arccos(-c_{\mathrm{p}})},  \\
j_{\mathrm{d}}&\equiv-\frac{\alpha k_{\mathrm{H}}}{1+\alpha\xi_{\mathrm{CPP}}}\cdot\frac{\sqrt{1-c_{\mathrm{p}}^2}}{b_{\mathrm{p}}\arccos(c_{\mathrm{p}})}.
\end{split}
\end{equation}
For the wall width of this branch, first its existence demands that when $n$ is even (odd),
$J_e/J_{\mathrm{p}}<j_{\Delta \mathrm{u}}$ ($J_e/J_{\mathrm{p}}>j_{\Delta \mathrm{d}}$) with
\begin{equation}\label{jDelta_down_jDleta_up_definitions}
\begin{split}
j_{\Delta\mathrm{u}}&\equiv\frac{k_{\mathrm{E}}}{\xi_{\mathrm{CPP}}}\cdot\frac{2 c_{\mathrm{p}}}{b_{\mathrm{p}}}\cdot\left(\frac{\pi^2}{4}-\arccos^2 c_{\mathrm{p}}\right)^{-1},  \\
j_{\Delta\mathrm{d}}&\equiv-\frac{k_{\mathrm{E}}}{\xi_{\mathrm{CPP}}}\cdot\frac{2 c_{\mathrm{p}}}{b_{\mathrm{p}}}\cdot\left[\arccos^2 (-c_{\mathrm{p}})-\frac{\pi^2}{4}\right]^{-1}.
\end{split}
\end{equation}
Since $\alpha\ll 1$ and $\xi_{\mathrm{CPP}} \ll 1$, $|j_{\Delta\mathrm{u(d)}}|\gg |j_{\mathrm{u(d)}}|$
and is usually out of experimental accessibility.
Thus only $j_{\mathrm{u(d)}}$ is considered when dealing with stability issue.
For the variation $\Delta=\Delta(\varphi_0)+\delta\Delta$, Eq. (\ref{Dynamical_equation_pt_polarizer}c) provides
\begin{equation}\label{Stability_analysis_Delta_pt_polarizer}
\frac{\pi^2\alpha}{6\gamma_0 M_s}\frac{\partial(\ln\delta\Delta)}{\partial t}=-\frac{2l_0^2}{\Delta^2(\varphi_0)},
\end{equation}
implying a stable wall width of this solution branch (see violet solid lines in Fig. \ref{fig2}).

Next we turn to the branch in Eq. (\ref{Solution_branch_2_pt_polarizer}).
By rewriting the first equation as $J_e/J_{\mathrm{p}}=-\alpha k_{\mathrm{H}}\cos\varphi'_0(1-c_{\mathrm{p}}^2\cos^2\varphi'_0)^{1/2}/[(1+\alpha\xi_{\mathrm{CPP}})b_{\mathrm{p}}\arccos(c_{\mathrm{p}}\cos\varphi'_0)]$ and analyzing its monotonicity,
the permitted current density range of this branch can be obtained.
Note that $J_e(\varphi'_0)=J_e(2\pi-\varphi'_0)$, we then focus on $\varphi'_0\in[0,\pi]$ thus $\sin\varphi'_0\ge 0$.
After standard calculus, one has
\begin{equation}\label{dj_dphi0}
\frac{\mathrm{d}}{\mathrm{d}\varphi'_0}\left(\frac{J_e}{J_{\mathrm{p}}}\right)=\frac{\alpha k_{\mathrm{H}}\sin\varphi'_0}{(1+\alpha\xi_{\mathrm{CPP}})b_{\mathrm{p}}}\cdot\frac{f(\zeta)}{\sqrt{1-\zeta^2}\cdot\arccos^2\zeta},
\end{equation}
with
\begin{equation}\label{fzeta_definition}
f(\zeta)=\left(1-2\zeta^2\right)\arccos\zeta+\zeta\sqrt{1-\zeta^2},\;\; \zeta\equiv c_{\mathrm{p}}\cos\varphi'_0.
\end{equation}
On the other hand, the counterpart of Eq. (\ref{Stability_analysis_phi_pt_polarizer}) for this solution
branch is
\begin{equation}\label{Stability_analysis_phiprime_pt_polarizer}
\frac{\partial(\ln\delta\varphi')}{\partial t}=\frac{\alpha\gamma_0 M_s k_{\mathrm{H}}\sin^2\varphi'_0}{(1+\alpha^2)(1-\zeta^2)\arccos\zeta}f(\zeta),
\end{equation}
The monotonicity analysis on $f(\zeta)$ provides us a critical value
$\zeta_0=-0.6256$ ($\Leftrightarrow P_0=0.3704$)\cite{He_EPJB_2013}.
When $c_{\mathrm{p}}<\zeta_0$ ($\Leftrightarrow P<P_0$), $f(\zeta)>0$.
This fact has two consequences: from Eq. (\ref{dj_dphi0}), $J_e/J_{\mathrm{p}}$ is
an increasing function on $\varphi'_0\in[0,\pi]$ thus acquires its minimum ($j_{\mathrm{d}}$)
at $\varphi'_0=0$ and maximum ($j_{\mathrm{u}}$) at $\varphi'_0=\pi$ (see Fig. \ref{fig2}(a)).
However, Eq. (\ref{Stability_analysis_phiprime_pt_polarizer}) tells us that now
this whole branch remains unstable thus is not physically preferred.
When $c_{\mathrm{p}}>\zeta_0$ ($\Leftrightarrow P>P_0$), $f(\zeta)$ first increases when $\varphi'_0$
runs from 0 to $\arccos(\zeta_0/c_{\mathrm{p}})$ and then decreases when
$\varphi'_0$ exceeds $\arccos(\zeta_0/c_{\mathrm{p}})$ to $\pi$.
Correspondingly, $J_e/J_{\mathrm{p}}$ increases from $j_{\mathrm{d}}$ to $j_{\mathrm{M}}=0.2172\alpha k_{\mathrm{H}}/[(1+\alpha\xi_{\mathrm{CPP}})b_{\mathrm{p}}c_{\mathrm{p}}]$ and
then decreases to $j_{\mathrm{u}}$, as illustrated in Fig. \ref{fig2}(b).
Meantime, from Eq. (\ref{Stability_analysis_phiprime_pt_polarizer}) only when
$\arccos(\zeta_0/c_{\mathrm{p}})<\varphi'_0<2\pi-\arccos(\zeta_0/c_{\mathrm{p}})$
the solution branch in Eq. (\ref{Solution_branch_2_pt_polarizer}) is stable,
which has been marked by red curves in Fig. \ref{fig2}(b).

Now we explain what happens physically when the CPP current density $J_e$ increases from 0 to large positive value.
If the wall initially lies in easy $xz-$plane with $\varphi|_{t=0}=0$, i.e. the magnetization at
wall center is parallel to the polarizer, then it always stays in this state with zero velocity no matter how large $J_e$ is.
While if the wall initially lies with $\varphi|_{t=0}=\pi$, i.e. the magnetization at
wall center is anti-parallel to the polarizer, it keeps on staying in this state until $J_e/J_{\mathrm{p}}$
increases to $j_{\mathrm{u}}$.
When $J_e$ is further enhanced a little bit, something interesting happens.
When the polarizer is not strong enough ($P<P_0$), the wall ``jumps"
to $\varphi=0$ state (through $\pi\rightarrow 0$ or $\pi\rightarrow 2\pi$ route depending on the
nature of external disturbances) and then keeps still.
On the contrary, if the polarizer is strong enough ($P_0<P\le 1$), the wall will evolve
into one of the two stable parts of the solution branch in Eq. (\ref{Solution_branch_2_pt_polarizer}).
Likely, which one it runs into is determined by the nature of external disturbances.
As $J_e/J_{\mathrm{p}}$ increases from $j_{\mathrm{u}}$ to $j_{\mathrm{M}}$, the wall
acquires a finite velocity as shown by the second equation of Eq. (\ref{Solution_branch_2_pt_polarizer}).
When $J_e/J_{\mathrm{p}}$ exceeds $j_{\mathrm{M}}$, the wall jumps to its nearest static branch
under external disturbance and then keeps in this state.

\begin{figure} [htbp]
	\centering
	\scalebox{0.41}[0.41]{\includegraphics[angle=0]{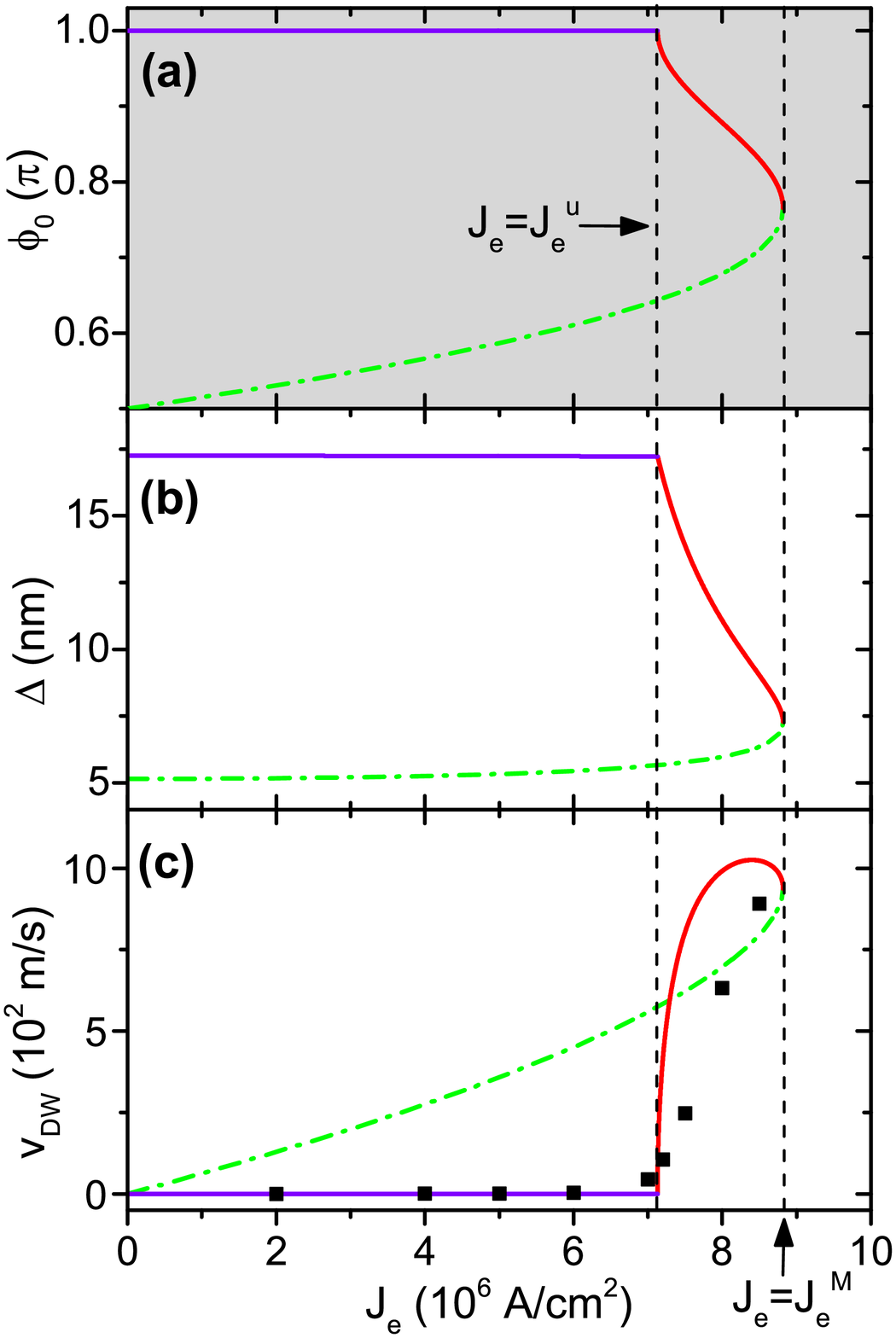}}
	\caption{(Color online) Dependence of the tilting angle (a), width (b) and velocity (c) of a TDW with $\eta=-1$ 
		on current density in a LNSV with CPP configuration and planar-transverse polarizer ($\mathbf{m}_{\mathrm{p}}=\mathbf{e}_x$).
		The subfigure (a) corresponds to the shaded area in Fig. \ref{fig2}(b).
		The free FM layer has the geometry of
		$50\times 3\times 8000$ $\mathrm{nm}^3$, with $M_s=1400$ kA/m, $J=30\times 10^{-12}$ J/m and $\alpha=0.007$.
		In addition, $P=0.6$ and $\xi_{\mathrm{CPP}}=0.1$ for spin-transfer process.
		The violet solid lines are the stable static branch in Eq. (\ref{Solution_branch_1_pt_polarizer}). The red solid (green dash-dot) curves comes from the stable (unstable) part of the finite-velocity branch in
		Eq. (\ref{Solution_branch_2_pt_polarizer}). Solid squares comes from OOMMF simulations.}\label{fig3}
\end{figure}

Next we do some numerical estimations.
The following magnetic parameters for Co are adopted (same as those in Ref. \cite{Khvalkovskiy_PRL_2009}):
$M_s=1400$ kA/m, $J=30\times 10^{-12}$ J/m, $\alpha=0.007$ and $\xi_{\mathrm{CPP}}=0.1$.
Thus the exchange length $l_0=4.94$ nm.
The geometry of free layer is $3\times 50\times 8000$ $\mathrm{nm}^3$, resulting
in three average demagnetization factors: $D_y=0.917251$, $D_x=0.082269$ and $D_z=0.000480$.
The crystalline anisotropy and edge roughness are both neglected, thus
$k_{\mathrm{E}}=D_x-D_z=0.081789$ and $k_{\mathrm{H}}=D_y-D_x=0.834982$.
Then $\Delta_0=l_0/\sqrt{k_{\mathrm{E}}}=17.3$ nm.
As indicated, to obtain stable propagating walls the spin polarization $P$ should satisfy $P>P_0=0.3704$.
Here we take $P=0.6$ as an example.
Then $b_{\mathrm{p}}=0.3832$ and $b_{\mathrm{p}}=0.8442$,
thus the extremal point is $\varphi_0^{\mathrm{M}}=\arccos(\zeta_0/c_{\mathrm{p}})=0.7657\pi$.
The upper limit of the current density for the stable static branch in Eq. (\ref{Solution_branch_1_pt_polarizer})
is $J_e^{\mathrm{u}}=j_{\mathrm{u}}\cdot J_{\mathrm{p}}=7.13\times 10^6$ $\mathrm{A/cm^2}$.
Meantime, the upper limit of the current density for the stable finite-velocity branch in Eq. (\ref{Solution_branch_2_pt_polarizer})
is $J_e^{\mathrm{M}}=j_{\mathrm{M}}\cdot J_{\mathrm{p}}=8.82\times 10^6$ $\mathrm{A/cm^2}$.
These two values are both not high for real applications.
Then the tilting angle, width and velocity of a TT ($\eta=-1$) TDW corresponding to the shaded 
area in Fig. \ref{fig2}(b) are calculated and plotted in Fig. \ref{fig3}.
We focus on the red curves which are the stable part of the finite-velocity branch in
Eq. (\ref{Solution_branch_2_pt_polarizer}).
Interestingly, at $J_e\approx 8.40\times 10^6$ $\mathrm{A/cm^2}$ the wall can
propagate along the LNSV at a velocity as high as 1025 m/s.
Therefore planar-transverse polarizers
have comparable current efficiency as perpendicular polarizers\cite{Khvalkovskiy_PRL_2009}.
To our knowledge, this has never been reported before in existing studies.

Another attracting quantity is the high ``differential mobility" ($\mathrm{d}v/\mathrm{d}J_e$)
around $J_e=J_e^{\mathrm{u}}$ ($\varphi'_0=\pi$), as shown by the red curve in Fig. \ref{fig3}(c).
From Eq. (\ref{dj_dphi0}), this infinity comes from the divergent behavior of 
$|\mathrm{d}\varphi'_0/\mathrm{d}J_e|\propto 1/|\sin\varphi'_0|\rightarrow +\infty$ at $J_e=J_e^{\mathrm{u}}$ ($\varphi'_0=\pi$).
Consequently, combining with Eq. (\ref{Solution_branch_2_pt_polarizer}),
we have $|\mathrm{d}v'_0/\mathrm{d} J_e| = |(\mathrm{d}v'_0/\mathrm{d}\varphi'_0)\cdot(\mathrm{d}\varphi'_0/\mathrm{d}J_e)|\propto |\cos 2\varphi'_0/\sin\varphi'_0|\rightarrow +\infty$.
This means that a slight increase of $J_e$ above $J_e^{\mathrm{u}}$ will lead to considerable
increase of wall velocity. 

\begin{figure} [htbp]
	\centering
	\includegraphics[width=0.45\textwidth]{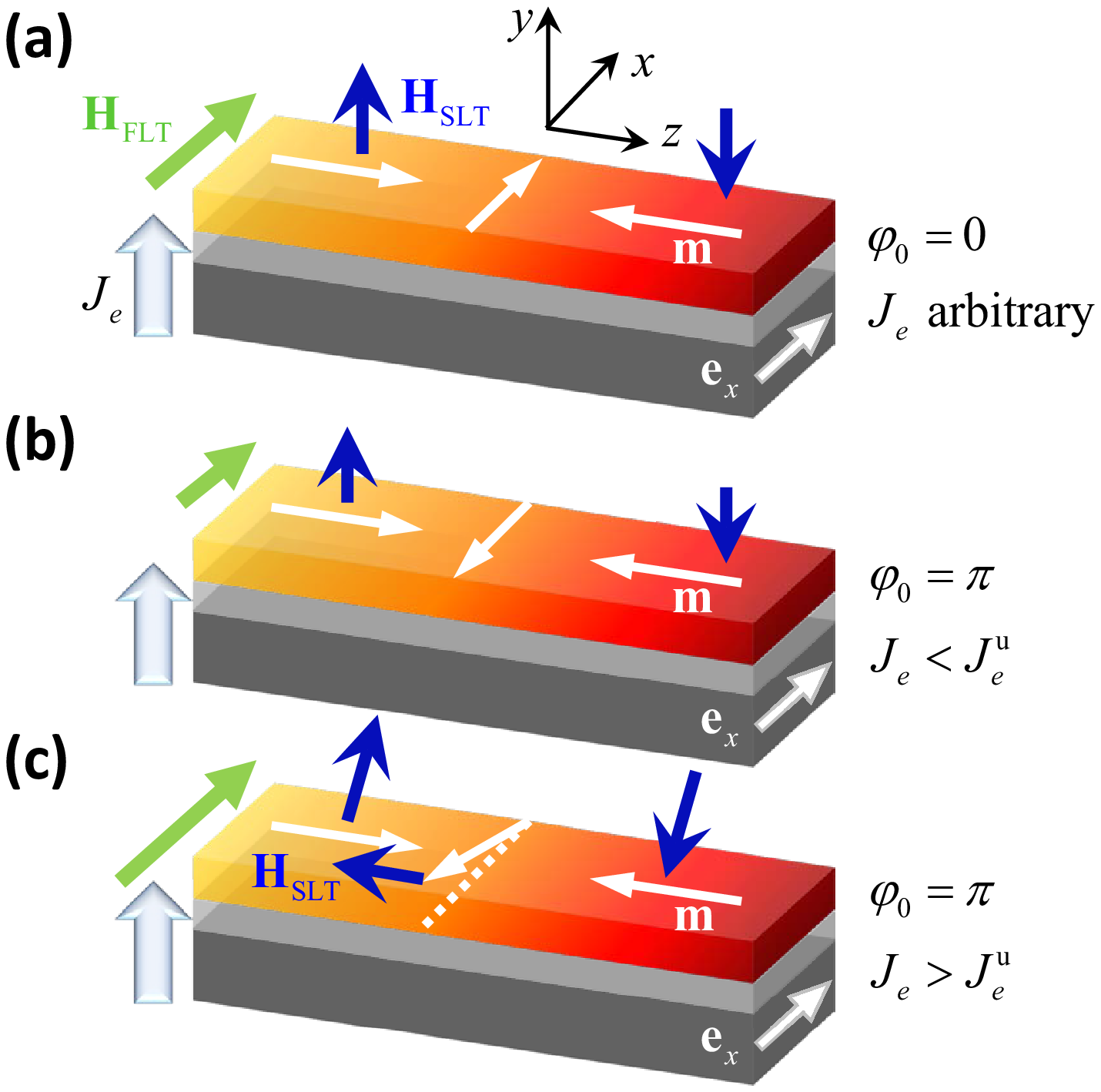}
	\caption{(Color online) Illustration of the physical mechanism responsible for the high differential 
		mobility around $J_e^{\mathrm{u}}$. (a) $\phi_0=0$ and $J_e$ is arbitrary. 
		(b) $\phi_0=\pi$ and $J_e<J_e^{\mathrm{u}}$. (c) $\phi_0=\pi$ and $J_e>J_e^{\mathrm{u}}$.
		In all sketches, $\eta=-1$. Green (blue) arrows represent $\mathbf{H}_{\mathrm{FLT}}$ ($\mathbf{H}_{\mathrm{SLT}}$). }\label{fig4}
\end{figure}

To make sure this high differential mobility around $J_e=J_e^{\mathrm{u}}$  
is a real effect rather than theoretical illusion,
we perform numerical simulations using the OOMMF package\cite{OOMMF} with the ``Xf\_STT" class embedded
which enables simulation on injection of multiple spin currents into a ferromagnet in OOMMF.
The resulting velocities are depicted by solid squares in Fig. \ref{fig3}(c).
The magnetic and geometric parameters are exactly the same with theoretical deductions.
In particular, the crystalline anisotropy and edge roughness are both neglected and
the demagnetization is turned on.
As preparation, a static TDW with $\eta=-1$ and $\phi'_0=\pi$ is generated at the wire center. 
Perpendicularly injected current density $-J_e\mathbf{e}_y$ manipulates the TDW dynamics.
First, a current density pulse with strength $3.0\times 10^7$ $\mathrm{A/cm^2}$
and duration 2.5 ns is applied to slightly push the wall away from its potential valley ($\varphi'_0=\pi$).
Then typical current densities ($\le 8.5\times 10^6$ $\mathrm{A/cm^2}$) are applied
and the wall velocities in stable traveling-wave mode are recorded.
Numerical data show that when $J_e<7.0\times 10^6$ $\mathrm{A/cm^2}$, the wall 
creeps forward for a distance and then stops.
When $J_e\ge 7.0\times 10^6$ $\mathrm{A/cm^2}$, the wall acquires high velocity very quickly as $J_e$ increases.
This critical current density is very close to the theoretical prediction $J_e^{\mathrm{u}}=7.13\times 10^6$ $\mathrm{A/cm^2}$.
At $J_e=8.5\times 10^6$ $\mathrm{A/cm^2}$, the wall velocity is around 900 m/s which is
comparable with the theoretical maximum (1025 m/s at $8.4\times 10^6$ $\mathrm{A/cm^2}$).
The difference between theoretical curve and simulation data comes from the fact
that around $J_e^{\mathrm{u}}$ the half wire with limited length (4 $\mathrm{\mu}$m)
is not enough for the wall to converge to its stable solution (\ref{Solution_branch_2_pt_polarizer}).

In fact, this large differential mobility can be understood physically.
By putting Eqs. (\ref{Lagrangian_density}) and (\ref{Dissipation_density_HePB})
into the generalized Eular-Lagrangian equation (\ref{Eular_Lagrangian_equation}) with $X=\theta(\phi)$, we obtain
the familiar Landau-Lifshitz-Gilbert (LLG) equation
\begin{equation}\label{LLG_vector}
\begin{split}
\frac{\partial \mathbf{m}}{\partial t}=&-\gamma_0\mathbf{m}\times\mathbf{H}^0_{\mathrm{eff}}+\alpha\mathbf{m}\times\frac{\partial \mathbf{m}}{\partial t}  \\
& \quad -\gamma_0 a_J\mathbf{m}\times(\mathbf{m}\times\mathbf{m}_{\mathrm{p}})-\gamma_0 b_J \mathbf{m}\times\mathbf{m}_{\mathrm{p}},
\end{split}
\end{equation}
where $a_J=\hbar J_e g/(2\mu_0 d e M_s)$, $b_J=\xi_{\mathrm{CPP}} a_J$, and $\mathbf{H}^0_{\mathrm{eff}}=-(\mu_0 M_s)^{-1}\delta\mathcal{E}_0/\delta\mathbf{m}$.
We denote the two effective fields related to SLT and FLT as $\mathbf{H}_{\mathrm{SLT}}=a_J(\mathbf{m}\times\mathbf{m}_{\mathrm{p}})$ and $\mathbf{H}_{\mathrm{FLT}}=b_J\mathbf{m}_{\mathrm{p}}$, respectively.
Note that Eq. (\ref{LLG_vector}) describes a gyrational magnetization dynamics accompanied by
a damping-induced motion towards the effective field.
For planar-transverse polarizers ($\mathbf{m}_{\mathrm{p}}=\mathbf{e}_x$), 
$\mathbf{H}_{\mathrm{FLT}}$ is always a uniform transverse field directed along $+\mathbf{e}_x$ thus can not induce 
TDW motion along $\mathbf{e}_z$.
However, it breaks the two-fold symmetry in $x-$direction: TDWs lying in $\varphi_0=0$ plane are always 
stable while at some critical current density ($J_e^{\mathrm{u}}$) TDWs initially lying in $\varphi_0=\pi$ plane 
will climb out of this potential valley formed by finite hard anisotropy in $y-$direction.

When $J_e<J_e^{\mathrm{u}}$, TDWs are still lying in $\varphi_0=n\pi$ valleys. 
Thus $\mathbf{H}_{\mathrm{SLT}}$ is perpendicularly to $\varphi_0=n\pi$ planes and directed oppositely about 
the wall center [see Fig. \ref{fig4}(a) and \ref{fig4}(b)].
The gyration around $\mathbf{H}_{\mathrm{SLT}}$ leads to temporary wall displacement.
At the same time, the damping process results in the tilting of magnetization towards $\mathbf{H}_{\mathrm{SLT}}$.
Correspondingly, magnetic charges appear at the opposite sides of the free layer and thus generate
a magnetostatic field that balances $\mathbf{H}_{\mathrm{SLT}}$. As a result, the wall stops and becomes static.

For TDWs initially lying in $\varphi_0=\pi$ valley and $J_e$ slightly exceeds $J_e^{\mathrm{u}}$, 
due to the symmetry about $\varphi_0=\pi$ plane, the magnetization at wall center departs from it randomly.
By denoting the new stable azimuthal angle as $\varphi'_0$ and from the famous ``Stoner-Wohlfarth asteroid" theorem\cite{Tannous_2008}, 
at critical point one has $H_{\mathrm{FLT}}\propto\cos\varphi'^3_0$ which leads to 
$|\mathrm{d}\varphi'_0/\mathrm{d}J_e|\propto|\mathrm{d}\varphi'_0/\mathrm{d}H_{\mathrm{FLT}}|\propto 1/|\sin\varphi'_0|\gg 1$.
This explains the high differential mobility around $J_e^{\mathrm{u}}$.
Now we take $0<\varphi'_0<\pi$ as an example [see Fig. \ref{fig4}(c)], in wall region $\mathbf{H}_{\mathrm{SLT}}$ has 
$-\mathbf{e}_z$ component.
For $\eta=-1$, this leads to a finite velocity along $+\mathbf{e}_z$ which explains 
the stable branch in Eq. (\ref{Solution_branch_2_pt_polarizer}).
When current density is too large ($>J_e^{\mathrm{M}}$), the generalized Walker
profile will collapse due to the anti-directed $\mathbf{H}_{\mathrm{SLT}}$ on the two sides of TDW
and vortex/antivortex may emerge which is out of the scope of this work.

In summary, dynamical behaviors of TDWs under planar-transverse polarizers in LNSVs with CPP configuration 
are quite different from known results in two aspects. 
First, in all well-investigated current-driven stack setups, including FM monolayers (CIP), 
FM/heavy-metal bilayers (CIP) and LNSVs with parallel and perpendicular polarizers (CPP), 
TDWs have a finite mobility in the entire range of current density when dealing with 
a sufficiently smooth and even sample (absence of intrinsic pinning due to imperfectness). 
This means TDWs will acquire a steady motion with finite velocity under finite charge current density, 
no matter how small the latter is. 
However in LNSVs with strong enough planar-transverse polarizers, steady wall motion with finite velocity 
can only occur when driving current exceeds a finite threshold of density. 
Second, at the onset of wall excitation, the differential mobility is very high due to the sudden change in 
steady tilting angle of TDWs as current density exceeds its lower limit a little bit. 
This allows TDWs to acquire high velocities under small current densities. 
The resulting current efficiency is comparable with that of perpendicular polarizers. 
When the current density exceeds its upper limit, TDWs jump to their nearest static branch. 
These two exotic behaviors should open new possibilities for developing magnetic nanodevices 
based on TDW propagation with low energy consumption: 
(a) When polarizers of LNSVs are made of 
magnetic materials with in-plane rather than perpendicular magnetic anisotropy, 
high current efficiency is still achievable as long as they are made planar-transverse.
(b) The high differential mobility around $J_e^{\mathrm{u}}$ makes these LNSVs candidates 
for high-sensitivity switches, etc.

\section{IV. DW dynamics under parallel and perpendicular polarizers}
The simulation work by Khvalkovskiy \textit{et. al.} proposed the high current efficiency
in LNSVs under parallel and perpendicular polarizers with ``$\mathbf{m}\cdot\mathbf{m}_{\mathrm{p}}$-independent"
STT coefficients\cite{Khvalkovskiy_PRL_2009}. 
Except for numerics, they also provided a 1D analysis for parallel polarizers in which the wall velocity and
tilting angle are two collective coordinates. 
However, for perpendicular polarizers the corresponding 1D analysis is absent.
Meantime, their simulations revealed that under perpendicular (parallel) polarizers
pure SLT (FLT) induces persistent wall displacement while pure FLT (SLT) does not. 
Therefore they conjectured that at low currents the large difference for the wall velocities
between perpendicular and planar polarizers is related to the factor $\xi_{\mathrm{CPP}}$
between the torques. However, the exact ratio of mobilities for these two cases under low currents
is not provided.
In this section, we perform systematic Lagrangian analysis and provide answers to these issues.

\subsection{IV.A Modified Lagrangian and dynamical equations}
For $\mathbf{m}\cdot\mathbf{m}_{\mathrm{p}}$-independent STT coefficients, the energy density
functional turns to
\begin{equation}\label{magnetic_energy_density_total_new}
\widetilde{\mathcal{E}}[\mathbf{m}]=\mathcal{E}_0[\mathbf{m}]-\mu_0 M_s \tilde{b}_J p_{\mathbf{m}},
\end{equation}
and the dissipation functional becomes
\begin{equation}\label{Dissipation_density_new}
\frac{\widetilde{\mathcal{F}}}{\mu_0 M_s^2}=\frac{\alpha}{2}\frac{\dot{\theta}^2+\dot{\phi}^2\sin^2\theta}{\gamma_0 M_s}- \frac{\tilde{a}_J}{M_s} (p_{\theta}\sin\theta\dot{\phi}-p_{\phi}\dot{\theta}),
\end{equation}
where $\tilde{a}_J=\hbar J_e P/(2\mu_0 d e M_s)$ and $\tilde{b}_J=\xi_{\mathrm{CPP}} \tilde{a}_J$.
Still, the generalized Walker profile is taken as
the configuration space of walls. After putting the wall center position $q(t)$, tilting angle
$\varphi(t)$ and width $\Delta(t)$ into Eq. (\ref{Eular_Lagrangian_equation}) successively,
and integrating over $z\in(-\infty,+\infty)$, a new set of dynamical equations are obtained
\begin{subequations}\label{Dynamical_equation_original_new}
	\begin{align}
	\alpha\eta\frac{\dot{q}}{\Delta}+\dot{\varphi}&=\gamma_0\left(\frac{\pi}{2}\tilde{a}_J p_{\varphi}+\tilde{b}_J \cos\theta_{\mathrm{p}}\right), \\
	\eta\frac{\dot{q}}{\Delta}-\alpha\dot{\varphi}&=\gamma_0 M_s k_{\mathrm{H}}\sin\varphi\cos\varphi   \nonumber \\
	    & \qquad +\gamma_0\left(\tilde{a}_J \cos\theta_{\mathrm{p}}-\frac{\pi}{2}\tilde{b}_J p_{\varphi}\right), \\
	\frac{\pi^2\alpha}{6}\frac{\dot{\Delta}}{\Delta}&=\gamma_0 M_s\left(\frac{l_0^2}{\Delta^2}-k_{\mathrm{E}}-k_{\mathrm{H}}\sin^2\varphi\right)   \nonumber  \\
	    & \qquad +\gamma_0\pi \tilde{b}_J \sin\theta_{\mathrm{p}}\cos(\varphi-\phi_{\mathrm{p}}).
	\end{align}
\end{subequations}

\subsection{IV.B Parallel polarizers}
For systematicness, we first briefly revisit TDW dynamics under parallel polarizers.
In this case, $\mathbf{m}_{\mathrm{p}}=\mathbf{e}_z$, thus $\theta_{\mathrm{p}}=0$ and then $p_{\varphi}=0$. 
The dynamical equations turn to
\begin{subequations}\label{Dynamical_equation_parallel_polarizer_new}
	\begin{align}
	\frac{1+\alpha^2}{\gamma_0}\frac{\eta\dot{q}}{\Delta}&=\frac{k_{\mathrm{H}}M_s}{2}\sin 2\varphi+(\tilde{a}_J+\alpha \tilde{b}_J), \\
	\frac{1+\alpha^2}{\gamma_0}\dot{\varphi}&=-\frac{\alpha k_{\mathrm{H}}M_s}{2}\sin 2\varphi+(\tilde{b}_J-\alpha \tilde{a}_J), \\
	\frac{\pi^2\alpha}{6\gamma_0 M_s}\frac{\dot{\Delta}}{\Delta}&=\frac{l_0^2}{\Delta^2}-k_{\mathrm{E}}-k_{\mathrm{H}}\sin^2\varphi.
	\end{align}
\end{subequations}
The first two equations reproduce Eq. (4) in Khvalkovskiy's work (see Ref. \cite{Khvalkovskiy_PRL_2009})
and the third one provides the TDW width.
For traveling-wave mode of TDW, $\dot{\varphi}=0$ and $\dot{\Delta}=0$.
This leads to a FLT-determined steady wall velocity
\begin{equation}\label{Solution_branch_parallel_polarizer}
\begin{split}
v_0&=\frac{\eta\Delta(\varphi_0)\gamma_0\xi_{\mathrm{CPP}} \tilde{a}_J}{\alpha},   \\
\sin 2\varphi_0&=\frac{2(\xi_{\mathrm{CPP}}-\alpha)\tilde{a}_J}{\alpha M_s k_{\mathrm{H}}},   \\
\Delta_1(\varphi_0)&=l_0(k_{\mathrm{E}}+k_{\mathrm{H}}\sin^2\varphi_0)^{-1/2}.
\end{split}
\end{equation}
For variation of $\varphi_0$, we have
\begin{equation}\label{Stability_analysis_phi_parallel_polarizer}
\frac{\partial(\ln\delta\varphi_0)}{\partial t}=-\frac{\alpha\gamma_0 M_s k_{\mathrm{H}}\cos 2\varphi_0}{2(1+\alpha^2)}.
\end{equation}
When $\cos 2\varphi_0>0$, i.e. $|\varphi_0-n\pi|<\pi/4$, the $\varphi_0-$solution is stable.
On the other hand, for variation of $\Delta_1$, one has
\begin{equation}\label{Stability_analysis_Delta_parallel_polarizer}
\frac{\pi^2\alpha}{6\gamma_0 M_s}\frac{\partial(\delta\Delta)}{\partial t}=-\frac{2l_0^2}{\Delta_1^2(\varphi_0)}\delta\Delta-\Delta_1(\varphi_0)k_{\mathrm{H}}\sin 2\varphi_0\delta\varphi_0.
\end{equation}
Thus the wall width should be stable as long as $\varphi_0$ is stable.

\begin{figure} [htbp]
	\centering
	\scalebox{0.41}[0.41]{\includegraphics[angle=0]{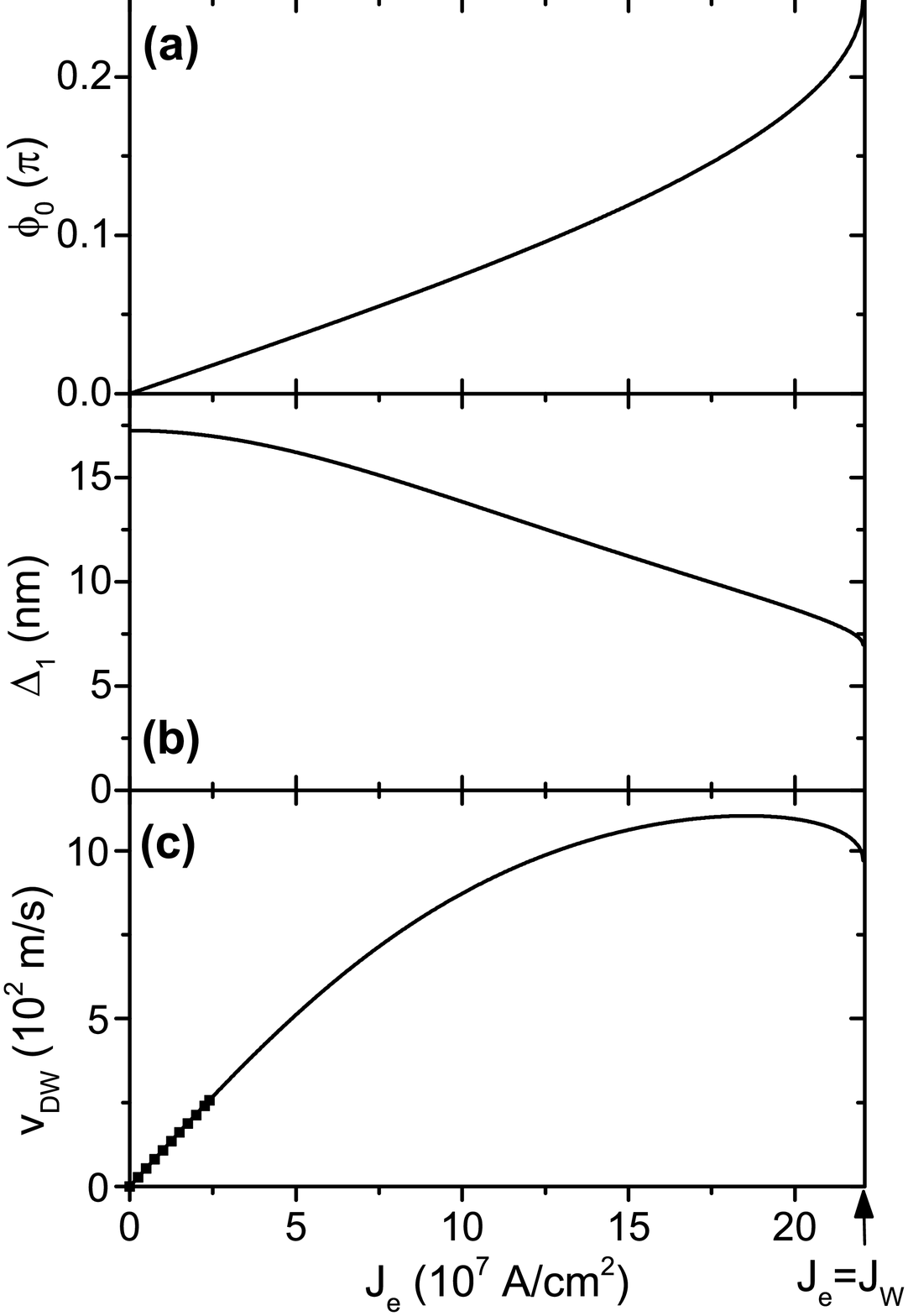}}
	\caption{Dependence of the tilting angle (a), width (b) and velocity (c) of a TDW on current density
		in a LNSV with CPP configuration and parallel polarizer ($\mathbf{m}_{\mathrm{p}}=\mathbf{e}_z$).
		The geometry and magnetic parameters of the free layer are the same as those in Fig. \ref{fig3}, except that
		the spin polarization is changed to $P=0.32$.
		The solid curves are from Eq. (\ref{Solution_branch_parallel_polarizer})
		and the solid squares in (c) are from Fig. 1(b) in Ref. \cite{Khvalkovskiy_PRL_2009}
		with exactly the same geometric and magnetic parameters. }\label{fig5}
\end{figure}

Next we compare our analytics with existing simulation data.
The geometry and magnetic parameters of the free layer are the same as those in the end of Sec. II, except that
the spin polarization is changed to $P=0.32$ (same as in Khvalkovskiy's work).
By requiring $|\sin 2\varphi_0|\le 1$, the Walker limit (under which traveling-wave mode survives) is
$J_{\mathrm{W}}=\alpha M_s k_{\mathrm{H}}/(2\kappa|\xi_{\mathrm{CPP}}-\alpha|)=2.20\times 10^8$ $\mathrm{A/cm^2}$.
However this is just theoretical prediction based on the generalized Walker profile.
Real simulations (see Fig. 1(b) of Ref. \cite{Khvalkovskiy_PRL_2009}) revealed that
TDWs disappear due to global-spin-transfer-induced domain excitation
when $J_e > 2.4\times 10^7$ $\mathrm{A/cm^2}$ which is an order of magnitude smaller $J_{\mathrm{W}}$.
Thus in traveling-wave mode, at most $\sin^2\varphi_0\sim 10^{-2}$ and
$\Delta_1(\varphi_0)\sim\Delta_0=17.3$ nm.
This leads to a constant wall mobility $\sim 1.09\times 10^{-5}$ $\mathrm{(m/s)/(A/cm^2)}$,
which perfectly explains the linear dependence of wall velocity on current density 
in Fig. 1(b) of Ref. \cite{Khvalkovskiy_PRL_2009}.
In Fig. \ref{fig5} of our work, analytical results from Eq. (\ref{Solution_branch_parallel_polarizer}) 
are plotted by solid curves.
Meantime, numerical data from Fig. 1(b) in Ref. \cite{Khvalkovskiy_PRL_2009} are indicated by solid squares.
Obviously as long as TDWs exist ($J_e < 2.4\times 10^7$ $\mathrm{A/cm^2}$), our theoretical results
are in good agreement with numerical simulations.

\subsection{IV.C Perpendicular polarizers}
Now $\mathbf{m}_{\mathrm{p}}=\mathbf{e}_y$, thus $\theta_{\mathrm{p}}=\pi/2$ and $\phi_{\mathrm{p}}=\pi/2$. 
Then $p_{\varphi}=\cos\varphi$ and Eq. (\ref{Dynamical_equation_original_new}) is simplified to
\begin{subequations}\label{Dynamical_equation_perp_polarizer_new}
	\begin{align}
	\frac{1+\alpha^2}{\gamma_0}\frac{\eta\dot{q}}{\Delta}&=\left[k_{\mathrm{H}}M_s\sin\varphi+\frac{\pi}{2}(\alpha \tilde{a}_J-\tilde{b}_J)\right]\cos\varphi, \\
	\frac{1+\alpha^2}{\gamma_0}\dot{\varphi}&=\left[\frac{\pi}{2}(\tilde{a}_J+\alpha \tilde{b}_J)-\alpha k_{\mathrm{H}}M_s\sin\varphi\right]\cos\varphi, \\
	\frac{\pi^2\alpha}{6\gamma_0 M_s}\frac{\dot{\Delta}}{\Delta}&=\left(\frac{l_0^2}{\Delta^2}-k_{\mathrm{E}}-k_{\mathrm{H}}\sin^2\varphi\right)+\frac{\pi \tilde{b}_J}{M_s}\sin\varphi.
	\end{align}
\end{subequations}
For steady traveling-wave mode, we need $\dot{\varphi}=0$ and $\dot{\Delta}=0$.
This leads to two branches of solution:
\begin{eqnarray}\label{Solution_branch_1_perp_polarizer_new}
&\varphi_0=\left(n+\frac{1}{2}\right)\pi,\quad v_0=0,&  \nonumber  \\ &\Delta_2(\varphi_0)=l_0\left[k_{\mathrm{E}}+k_{\mathrm{H}}-(-1)^n\frac{\pi \tilde{b}_J}{M_s}\right]^{-1/2},&
\end{eqnarray}
and
\begin{equation}\label{Solution_branch_2_perp_polarizer_new}
\begin{split}
\sin\varphi'_0&=\frac{\pi}{2} \frac{1+\alpha\xi_{\mathrm{CPP}}}{\alpha} \frac{\tilde{a}_J}{k_{\mathrm{H}}M_s},   \\
v'_0&=\frac{\pi}{2}\frac{\eta\Delta_2(\varphi'_0)\gamma_0 \tilde{a}_J}{\alpha}\cos\varphi'_0,    \\
\Delta_2(\varphi'_0)&=l_0\left(k_{\mathrm{E}}+k_{\mathrm{H}}\sin^2\varphi'_0-\frac{\pi \tilde{b}_J}{M_s}\sin\varphi'_0\right)^{-1/2}.
\end{split}
\end{equation}

Then we perform stability analysis to these two branches. 
For the one in Eq. (\ref{Solution_branch_1_perp_polarizer_new}),
after taking variation of $\varphi_0$ and substituting it into
Eq. (\ref{Dynamical_equation_perp_polarizer_new}b), one has
\begin{equation}\label{Stability_analysis_phi_perp_polarizer}
\frac{\partial(\ln\delta\varphi_0)}{\partial t}=-\frac{\alpha\gamma_0 M_s k_{\mathrm{H}}}{1+\alpha^2}\left[\frac{\pi}{2} \left(\frac{1}{\alpha}+\xi_{\mathrm{CPP}}\right) \frac{(-1)^n \tilde{a}_J}{k_{\mathrm{H}}M_s}-1\right].
\end{equation}
Then we define $J_1\equiv 4|\xi_{\mathrm{CPP}}-\alpha|\pi^{-1}(1+\alpha\xi_{\mathrm{CPP}})^{-1}J_{\mathrm{W}}$.
When $J_e>J_1$ ($n$ is even) or $J_e<-J_1$ ($n$ is odd),
$(-1)^n (\alpha^{-1}+\xi_{\mathrm{CPP}}) \tilde{a}_J\pi/(2k_{\mathrm{H}}M_s)-1>0$ always holds
thus the $\varphi_0=(n+1/2)\pi$ solution in the first branch is stable.
For the wall width of this branch, similar variational analysis provides the same result 
as in Eq. (\ref{Stability_analysis_Delta_pt_polarizer}), 
implying that the static solution at $\varphi_0=(n+1/2)\pi$ always has a stable wall width.

Then we move to the other branch in Eq. (\ref{Solution_branch_2_perp_polarizer_new}).
The solution $\varphi'_0$ requires $|\sin\varphi'_0|\le 1$, which is equivalent to $|J_e|\le J_1$.
After varying $\varphi'_0$ by $\delta\varphi'$ and putting into Eq. (\ref{Dynamical_equation_perp_polarizer_new}b),
we have
\begin{equation}\label{Stability_analysis_phiprime_perp_polarizer}
\frac{\partial(\ln\delta\varphi')}{\partial t}=-\frac{\alpha\gamma_0 M_s k_{\mathrm{H}}\cos^2\varphi'_0}{1+\alpha^2},
\end{equation}
implying that $\varphi'_0-$solution is always stable. The corresponding TDW velocity can be explicitly written out as
\begin{equation}\label{TDW_velocity_branch_2_perp_polarizer}
v'_0=\frac{\pi}{2}\frac{\eta\Delta_2(\varphi'_0)\gamma_0 \tilde{a}_J}{\alpha}(-1)^m\sqrt{1-\left(\frac{\pi}{2} \frac{1+\alpha\xi_{\mathrm{CPP}}}{\alpha k_{\mathrm{H}}M_s} \tilde{a}_J\right)^2},
\end{equation}
in which ``$(-1)^m$" comes from the initial condition ($\varphi'_0|_{t=0}=m\pi$ at $t=0$).
For $|J_e|\ll J_1$, one has
\begin{equation}\label{TDW_velocity_appro_branch_2_perp_polarizer}
v'_0\approx\frac{\pi}{2}\frac{\eta\Delta_0\gamma_0 \tilde{a}_J}{\alpha}(-1)^m.
\end{equation}
Clearly it has a mobility larger than that of ``parallel-polarizer" case [see Eq. (\ref{Solution_branch_parallel_polarizer})] 
by a factor of $\pi/(2\xi_{\mathrm{CPP}})\approx 15.7$, thus well explains the higher current efficiency of
perpendicular polarizers.
When $|J_e|\rightarrow J_1$, the $\varphi'_0-$solution converges to $\varphi_0-$branch
with zero wall velocity.

\begin{figure} [htbp]
	\centering
	\scalebox{0.41}[0.41]{\includegraphics[angle=0]{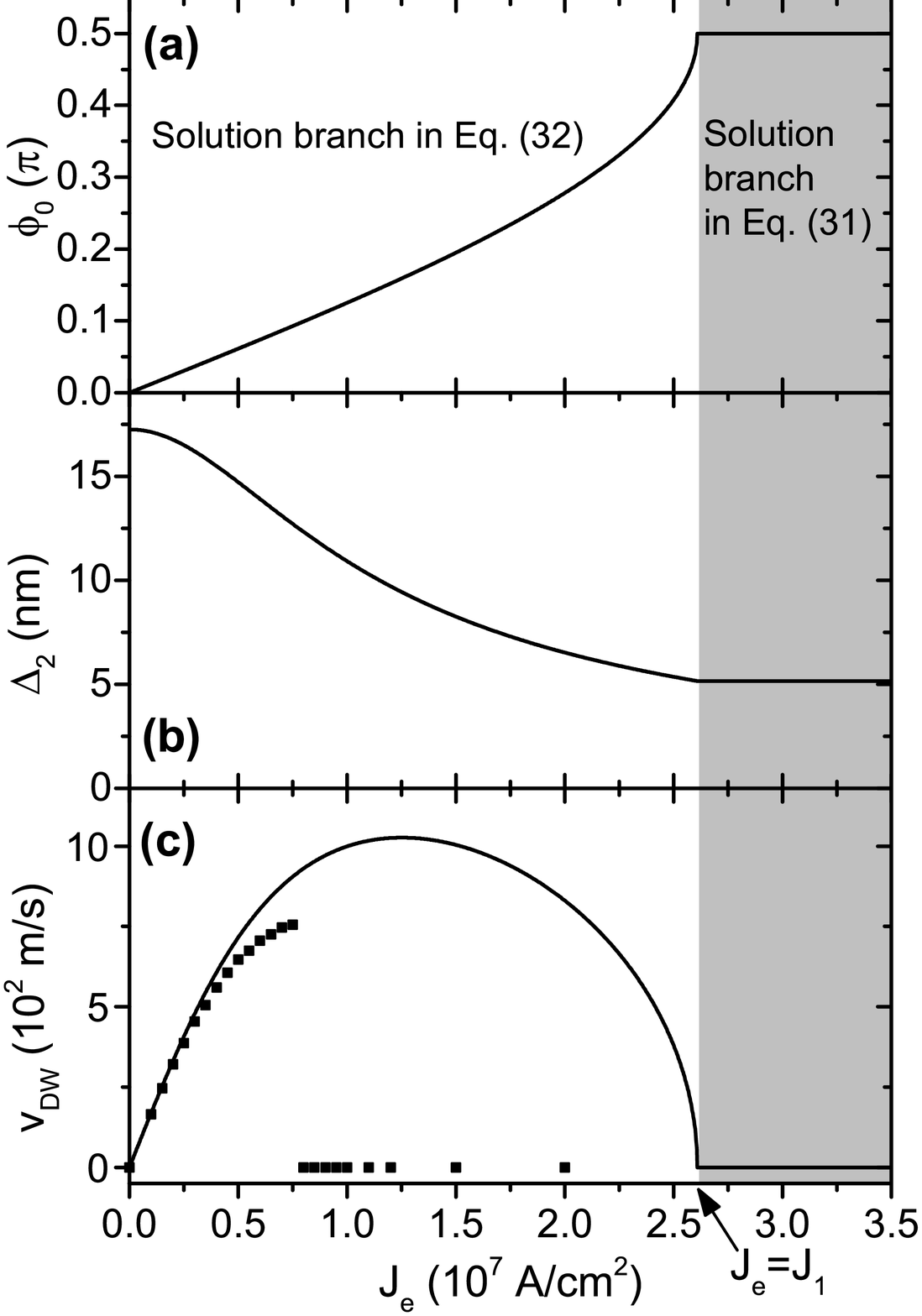}}
	\caption{Dependence of the tilting angle (a), width (b) and velocity (c) of a TDW on current density
		in a LNSV with CPP configuration and perpendicular polarizer ($\mathbf{m}_{\mathrm{p}}=\mathbf{e}_y$).
		The geometry and magnetic parameters of the free layer are the same as those in Fig. \ref{fig5}.
		The solid curves in the white-background area are the solution branch
		in Eq. (\ref{Solution_branch_2_perp_polarizer_new}) and the solid lines in the shaded area
		are those from Eq. (\ref{Solution_branch_1_perp_polarizer_new}).
		The solid squares in (c) are from Fig. 2(b) in Ref. \cite{Khvalkovskiy_PRL_2009}. }\label{fig6}
\end{figure}

For the same magnetic parameters as in parallel polarizers, $J_1=0.1183J_{\mathrm{W}}=2.61\times 10^7$ $\mathrm{A/cm^2}$.
For HH TDWs ($\eta=+1$) and standard initial condition ($\varphi_0|_{t=0}=\varphi'_0|_{t=0}=0$),
the solution branch in Eq. (\ref{Solution_branch_1_perp_polarizer_new})
[Eq. (\ref{Solution_branch_2_perp_polarizer_new})] is plotted in Fig. \ref{fig6} by solid line (curve)
in shaded (white-background) area.
In addition, simulation data from Fig. 2(b) in Ref. \cite{Khvalkovskiy_PRL_2009} are depicted 
in our Fig. \ref{fig6}(c) by solid squares.
Clearly when $J_e\le 0.3\times 10^7$ $\mathrm{A/cm^2}$, our analytics
coincides with simulation data very well. For larger current density, the wall
configuration in simulations will be distorted from the standard Walker profile due to
global spin transfers, thus leads to the inconsistency between analytics and simulations.

\subsection{IV.D Comparison with experimental data}
As mentioned in Sec. I, to our knowledge there are two groups of experimental work. 
In the first group, currents with definite density inject perpendicularly into LNSVs 
or half-ring MTJs (genuine CPP configurations). 
While in the second group, in-plane current flows through ZigZag LNSVs and 
the “vertical spin current” is suggested to be the source of DW velocity boosting, 
however the corresponding spin current density is hard to estimate. 
Therefore we focus on the first group in which genuine CPP configuration 
with definite current density is under investigation. 
Furthermore, our analytics is obtained in a strip geometry (resulting in three 
averaged demagnetization factors, and hence $k_{\mathrm{E}}$ and $k_{\mathrm{H}}$), 
thus can not directly apply to half-ring geometry. 
In summary, the best case to make the comparison is the first case in the first group, 
which is the experimental work by Boone \textit{et. al.} in 2010\cite{Boone_PRL_2010_exp}
on LNSVs with parallel polarizers.

In their work, the free layer is made of the nickel-rich nickel-iron alloy (NRNIA) with $M_s=430$ kA/m and
the crystalline anisotropy is neglected.
Its geometry ($3\times 90\times 5000$ $\mathrm{nm}^3$) provides three average demagnetization 
factors: $D_y=0.9473$, $D_x=0.05182$ and $D_z=0.00088$.
Thus we have $k_{\mathrm{E}}=D_x-D_z=0.0509$ and $k_{\mathrm{H}}=D_y-D_x=0.8955$.
The resulting coercive force is $k_{\mathrm{E}}M_s=275$ Oe, which is consistent with
experimental measurements (NRNIA reversal at $+200$ and $-300$ Oe) in Fig. 2(b) of Ref. \cite{Boone_PRL_2010_exp}.
The exchange stiffness ($J$) has not been explicitly provided.
However from the fixed wall width ($\lambda=53$ nm) they adopted in simulations,
we have $J=\mu_0 k_{\mathrm{E}} M_s^2\lambda^2/2=16.6\times 10^{-12}$ J/m.
Furthermore, the conversion coefficient from current density to SLT strength is
$\kappa=\tilde{a}_J/J_e=\hbar P/(2\mu_0 d e M_s)=1.32\times 10^{-3}$ $\mathrm{(A/m)/(A/cm^2)}$ for $P=0.65$.
Now we estimate the wall mobility under small driving currents 
where the wall width can be viewed as constant ($\lambda=53$ nm). 
Note that they obtained a damping coefficient from a fitting to the rectified voltage
with zero-FLT assumption. 
However as indicated by Khvalkovskiy \textit{et. al.}, FLT is crucial for 
TDW dynamics in LNSVs with parallel polarizers.
Therefore we adopt the typical NRNIA value $\alpha=0.01$ rather than their fitting parameter.
Moreover, we assume $\xi_{\mathrm{CPP}}=0.1$ which is the maximum permissible in Ref. \cite{Boone_PRL_2010_exp}.
From Eq. (\ref{Solution_branch_parallel_polarizer}), the wall mobility is
$|v_0/J_e|=\lambda\gamma_0\xi_{\mathrm{CPP}}\kappa/\alpha=1.55\times 10^{-4}$ $\mathrm{(m/s)/(A/cm^2)}$.
This agrees well with their experimental data for $J_e < 2\times 10^6$ $\mathrm{A/cm^2}$ in their Fig. 4(b).
On the other hand, the fitting result $\alpha=0.09$ leads to a wall mobility 
of $1.72\times 10^{-5}$ $\mathrm{(m/s)/(A/cm^2)}$.
This is an order of magnitude smaller than the experimental observations thus should be abandoned.

\section{V. Further boosting by UTMFs}
In real magnetic nondevices composed of LNSVs, to further boost TDWs' propagation, a UTMF
\begin{equation}\label{UTMF_vec}
\mathbf{H}_{\mathrm{TMF}}=H_{\perp}(\cos\Phi_{\perp},\sin\Phi_{\perp},0)
\end{equation}
can be applied, with $H_{\perp}$ and $\Phi_{\perp}$ being its strength and
orientation, respectively.
Meanwhile, the pinned layer is assumed to be unaffected which is a harmless simplification 
and will not affect our main conclusion.
Nevertheless, rigorous profile and velocity of TDWs under an arbitrary UTMF are hard to obtain
due to the mismatch between symmetries in different energy terms in transverse direction.
Since we focus on the traveling-mode at low current density,
the 1D-AEM\cite{Goussev_PRB_2013,Goussev_Royal_2013,jlu_PRB_2016,jlu_SciRep_2017,jlu_Nanomaterials_2019} 
on LLG equation shall provide useful information.
Recalling the results in Sec. III, for TDWs moving under planar-transverse polarizers, 
1D-AEM is not applicable since stable wall motion with finite velocity can only be excited 
for current density exceeding a finite threshold.
Hence in this section, we present the results for parallel and perpendicular polarizers.

\subsection{V.A Parallel polarizers}
The 1D-AEM needs static profiles of TDWs as the basis to calculate the response
of the system under external stimuli. 
Depending on UTMF strength, static TDWs take different profiles.
Therefore we discuss the ``small UTMF" and ``finite UTMF" cases separately.

For small UTMFs, the CCP current density, UTMF, and inverse of time are rescaled
simultaneously, that is $\tilde{a}_J=\epsilon \tilde{a}_J^0$, $\tilde{b}_J=\epsilon \tilde{b}_J^0$, $H_{\perp}=\epsilon h_{\perp}$ and $1/t=\epsilon(1/\tau)$,
where $\epsilon$ is the rescaling infinitesimal. The real solution of the LLG equation
is expanded as $\Omega(z,t) = \Omega_0(z,\tau)+\epsilon\Omega_1(z,\tau)+O(\epsilon^2)$ with $\Omega=\theta,\phi$.
Putting them back into the original LLG equation (\ref{LLG_vector}), the solution to the zeroth-order equation
is the Walker ansatz.
At the first order of $\epsilon$, with the help of zeroth-order solutions, the differential equation
about $\theta_1$ reads,
\begin{equation}\label{smallUTMF_CPPz_Ltheta1}
\begin{split}
F_{\mathrm{s}}&= \mathcal{L}\theta_1,\quad
\mathcal{L}\equiv \frac{2J}{\mu_0 M_s}\left(-\frac{\mathrm{d}^2}{\mathrm{d}z^2}+\frac{\theta'''_0}{\theta'_0}\right),  \\
F_{\mathrm{s}} &\equiv  \left[\frac{\eta\alpha (z_0)_{\tau}}{\gamma_0\Delta_0}-\tilde{b}_J^0\right]\sin\theta_0+(-1)^n h_{\perp}\cos\theta_0\cos\Phi_{\perp},
\end{split}
\end{equation}
where $(z_0)_{\tau}\equiv\mathrm{d} z_0/\mathrm{d}\tau$ and a prime means $\mathrm{d}/\mathrm{d}z$.
The subscript ``s" indicates
the ``small UTMF" case.
Note that $\mathcal{L}$ is the same 1D self-adjoint Schr\"{o}dinger operator as given in
Refs. \cite{Goussev_PRB_2013,Goussev_Royal_2013,jlu_PRB_2016,jlu_SciRep_2017,jlu_Nanomaterials_2019}.
Following the ``Fredholm alternative", by demanding $\theta'_0$ (kernel of $\mathcal{L}$)
to be orthogonal to the function $F_{\mathrm{s}}$
defined in Eq. (\ref{smallUTMF_CPPz_Ltheta1}),
TDW velocity in traveling-wave mode under small UTMFs is
\begin{equation}\label{smallUTMF_velocity_CPPz}
V_{\mathrm{s}}=\epsilon (z_0)_{\tau}=\eta\gamma_0\Delta_0 \tilde{b}_J/\alpha,
\end{equation}
which reproduces the rigorous result in Eq. (\ref{Solution_branch_parallel_polarizer}).

For finite UTMFs, we rescale the current density and the TDW velocity ($V_{\mathrm{f}}$) simultaneously,
i.e. $\tilde{a}_J=\epsilon \tilde{a}_J^0$, $\tilde{b}_J=\epsilon \tilde{b}_J^0$ and $V_{\mathrm{f}}=\epsilon v$ in which the subscript ``f" denotes the ``finite UTMF" case.
By introducing the traveling coordinate $\tilde{z}\equiv z-V_{\mathrm{f}} t=z-\epsilon v t$,
$\theta(z,t)$ and $\phi(z,t)$ are expanded as $\Omega(z,t) = \Omega_0(\tilde{z})+\epsilon\Omega_1(\tilde{z})+O(\epsilon^2)$
with $\Omega=\theta,\phi$.
Substituting them into the LLG equation,
an approximate polar angle profile $\theta_0$ (solution to the zeroth-order equations) of the wall is obtained
\begin{equation}\label{finiteTMF_theta0}
	\ln\frac{\sin\theta_0-\sin\theta_{\infty}}{1+\cos(\theta_0+\theta_{\infty})}=\frac{\eta\tilde{z}}{\Delta(\phi_{\infty})/\cos\theta_{\infty}},\quad
\end{equation}
with
\begin{equation}\label{finiteTMF_theta0_definitions}
\begin{split}
	\phi_{\infty}&=\tan^{-1}\left[k_{\mathrm{E}}\tan\Phi_{\perp}/(k_{\mathrm{E}}+k_{\mathrm{H}})\right], \\
	\theta_{\infty}&=\sin^{-1}\frac{H_{\perp}}{M_s\sqrt{k_{\mathrm{E}}^2\cos^2\phi_{\infty}+(k_{\mathrm{E}}+k_{\mathrm{H}})^2\sin^2\phi_{\infty}}}, \\
	\Delta(\phi_{\infty})&=l_0\left(k_{\mathrm{E}}+k_{\mathrm{H}}\sin^2\phi_{\infty}\right)^{-1/2},
\end{split}
\end{equation}
in which $\theta_{\infty}$ ($\phi_{\infty}$) is the polar (azimuthal) angle of magnetization in domains.
At the first order of $\epsilon$, after similar process as in field-driven case\cite{jlu_PRB_2016}, 
the equation about $\theta_1$ is
\begin{equation}\label{finiteTMF_CPPz_Ltheta1}
\mathcal{L}(\theta_1) = F_{\mathrm{f}}\equiv v\gamma_0^{-1}(\alpha\theta'_0-\sin\theta_0\phi'_0)-\tilde{b}_J^0\sin\theta_0.
\end{equation}
where a ``prime" means $\mathrm{d}/\mathrm{d}\tilde{z}$.
Again, $\theta'_0$ (kernel of $\mathcal{L}$) should be orthogonal to the function $F_{\mathrm{f}}$. 
After similar calculation, TDW velocity in traveling-wave mode under finite UTMF is,
\begin{equation}\label{finiteTMF_velocity_CPPz}
\begin{split}
V_{\mathrm{f}}&=u(\theta_{\infty})\frac{\eta\gamma_0\Delta(\phi_{\infty}) \tilde{b}_J}{\alpha},   \\
u(\theta_{\infty})&=\frac{2\cos\theta_{\infty}}{2\cos\theta_{\infty}-(\pi-2\theta_{\infty})\sin\theta_{\infty}}.
\end{split}
\end{equation}
This clearly shows that UTMFs can boost TDW propagation by a factor $u(\theta_{\infty})$, 
which has been well studied in Ref. \cite{jlu_PRB_2016}.

\subsection{V.B Perpendicular polarizers}
For small UTMFs, after similar rescaling, expansion and substitution operations,
the differential equation about $\theta_1$ is,
\begin{equation}\label{smallUTMF_CPPx_Ltheta1}
\mathcal{L}\theta_1 = \frac{\eta\alpha (z_0)_{\tau}\sin\theta_0}{\gamma_0\Delta_0} +(-1)^n \left(h_{\perp}\cos\theta_0\cos\Phi_{\perp}-\tilde{a}_J^0\right).
\end{equation}
The corresponding wall velocity is,
\begin{equation}\label{smallUTMF_velocity_CPPx}
V_{\mathrm{s}}=\epsilon (z_0)_{\tau}=(-1)^n\eta\pi\gamma_0\Delta_0 \tilde{a}_J/(2\alpha),
\end{equation}
which is the $\varphi'_0\rightarrow n\pi$ limit of Eq. (\ref{Solution_branch_2_perp_polarizer_new}).

For finite UTMFs, the equation about $\theta_1$ is
\begin{equation}\label{finiteTMF_CPPx_Ltheta1}
\mathcal{L}\theta_1 = \frac{v}{\gamma_0}(\alpha\theta'_0-\sin\theta_0\phi'_0) -\tilde{a}_J^0\cos\phi_0+\tilde{b}_J^0\cos\theta_0\sin\phi_0.
\end{equation}
The existence condition of $\theta_1-$solution provides
\begin{equation}\label{finiteTMF_velocity_CPPx}
\begin{split}
V_{\mathrm{f}} &\approx \omega(\theta_{\infty})\frac{\eta \gamma_0\Delta(\phi_{\infty})\tilde{a}_J}{\alpha}\cos\phi_{\infty},  \\
\omega(\theta_{\infty})&=\frac{\pi-2\theta_{\infty}}{2\cos\theta_{\infty}-(\pi-2\theta_{\infty})\sin\theta_{\infty}}.
\end{split}
\end{equation}
Simple calculus shows that $\omega(\theta_{\infty})$ has similar divergent behavior as $u(\theta_{\infty})$ 
when $H_{\perp}\rightarrow H_{\perp}^{\mathrm{max}}$, thus considerably boost TDW motion.
Interestingly, in LNSVs with perpendicular polarizers, TDW motion can be manipulated not only 
by UTMF strength (via ``$\omega(\theta_{\infty})$") but also its orientation (via ``$\cos\phi_{\infty}$"). 
This comes from the fact that polarized electrons always act as an extra time-dependent effective field in hard axis.
For TDWs with $\phi_{\infty}\neq n\pi$, magnetization in wall region rotates
around the effective field hence results in a translational wall displacement along ``$\eta\mathbf{e}_z$" direction.
Meanwhile, projection of SLT to the hard axis $\mathbf{e}_y$ contributes to ``$\cos\phi_{\infty}$". 
These lead to the final ``$\eta\cos\phi_{\infty}$" factor in Eq. (\ref{finiteTMF_velocity_CPPx}).

\section{\label{Section_Conclusion} VI. Summary}
In this work, TDW dynamics in LNSVs with CPP configurations are systematically investigated 
within Lagrangian framework. 
When STT coefficients take the Slonczewski's original form,
our results show that stable traveling-wave motion of TDWs with finite velocity can 
survive for strong enough planar-transverse polarizers, with the current efficiency
comparable with that of perpendicular ones. 
More importantly, TDWs have ultra-high differential mobility around the onset of stable wall excitation. 
These results should provide insights for developing magnetic nanodevices with low energy consumption.
For $\mathbf{m}\cdot\mathbf{m}_{\mathrm{p}}$-independent STT coefficients,
analytics for parallel and perpendicular polarizers perfectly explains existing simulations and experiments.
At last, further boosting of TDWs by external UTMFs are investigated with help of 1D-AEM and turns out to be efficient.


\section{Acknowledgement}
M. L. is supported by the Project of Hebei Province Higher Educational Science and Technology Program (QN2019309).
Z. A. is funded by the Hebei Province Department of Education (GCC2014025).
J. L. acknowledges the support from National Natural Science Foundation of China (Grant No. 11374088).


\end{document}